\documentclass[preprint,prd,noshowpacs]{revtex4}
\usepackage{amssymb}
\usepackage{amsmath}
\usepackage{graphicx}

\begin{document}
\title{Gravitons to Photons -- attenuation of gravitational waves
\footnote{Essay written for the Gravity Research Foundation 2015 Awards for Essays on Gravitation.}}
\author{Preston Jones}
\email{pjones@csufresno.edu}

\affiliation{Physics Department, California State University Fresno, Fresno, CA 93740\\
and Embry Riddle Aeronautical University, Prescott, AZ 86301}

\author{Douglas Singleton}
\email{dougs@csufresno.edu}

\affiliation{Physics Department, California State University Fresno, Fresno, CA 93740}

\date{\today}

\begin{abstract}
In this essay we examine the response of an Unruh-DeWitt detector (a quantum two-level system) to a gravitational
wave background. The spectrum of the Unruh-Dewitt detector is of the same form as some scattering processes or
three body decays such as muon-electron scattering or muon decay. Based on this similarity we propose that the Unruh-DeWitt
detector response implies a ``decay" or attenuation of gravitons, $G$, into photons, $\gamma$, via 
$G + G \rightarrow \gamma + \gamma $ or $G \rightarrow \gamma + \gamma + G$. Over large distances such a 
decay/attenuation may have consequences in regard to the detection of gravitational waves.
\end{abstract}

\maketitle

In the absence of an acceptable theory of quantum gravity the semi-classical theory of quantum fields in curved space-times 
\cite{BD,DeWitt75,Parker09} is the best available approach to the study of the interaction of quantum fields with gravitational 
phenomena. One of the key results of this semi-classical approach is the phenomenon of particle creation in a gravitational background. The usual example of this is Hawking radiation \cite{Hawking74} - the creation of photons in the background of a black hole. One of the most direct ways to check if field quanta/photons are created by a given gravitational background 
is through the use of an Unruh-DeWitt (UD) detector \cite{DeWitt75,Unruh76}. An UD detector is a quantum system with two 
energy levels. A simple, realistic model of an UD detector is an electron  placed in a uniform 
magnetic fields \cite{Bell83}. By placing the UD detector in a given gravitational background, and observing how the upper 
energy level of the quantum system is populated, one can determine the emission/absorption spectrum 
for a given space-time. The UD detector method can be used to find the radiation associated 
with several well-known space-times: (i) the thermal radiation, with Hawking temperature $T = \frac{\hbar c^3}{8 k_B \pi M G}$,  
of a black hole of mass $M$; (ii) the thermal radiation, with temperature $T= \frac{\hbar a}{2 \pi c k_B}$, observed by an 
eternally accelerating Rindler observer.

In this essay we want to examine the response of an UD detector to a gravitational plane wave background. The background 
of a gravitational plane wave is different from a black hole space-time in that the gravitational plane wave has no horizon. 
In this regard the response of an UD detector to a gravitational plane wave background is similar to the response of 
an UD detector undergoing circular motion as studied in \cite{Bell83,letaw,Akhmedov07,Akhmedov07a}. Neither the rotating 
space-time nor the gravitational wave space-time  has a horizon (although the circular rotating metric has a light
surface). The spectrum of the UD detector in the background of a plane gravitational wave is obtained and found to have 
the same form as some Standard Model scattering or three body decays.
Based on this similarity we propose that this represents, in Feynman diagram language, either a ``decay" of the 
initial two gravitons into two photons ({\it i.e.} $G + G \rightarrow \gamma + \gamma$) or the  attenuation of one initial
graviton into two photons plus a lower frequency/energy graviton ({\it i.e.} $G \rightarrow \gamma + \gamma + G$). 
Such processes, as well as related processes like graviton-photon ``Compton scattering" 
$G + \gamma \rightarrow  G + \gamma$, have been investigated in \cite{Kallberg06,Brodin06,Bohr08} using 
using Feynman  diagrammatic methods or other techniques. The attenuation process,
$G \rightarrow \gamma + \gamma + G$, is related to the previously studied graviton decay \cite{Mondanese95} process 
$G \rightarrow G+ G + G$. Both $G + G \rightarrow \gamma + \gamma$ and $G \rightarrow \gamma + \gamma + G$ would lead
to a weakening of the gravitational wave amplitude/graviton intensity with distance from the source.

In order to examine particle production from quantum vacuum fluctuations we consider the special case of a gravitational 
plane wave \cite{Taylor14} traveling in the $x_3= +z$ direction with diagonal metric. Taking the UD detector to be located 
as $z=0$ the metric at this point is $g_{00}  =  - 1,\;g_{11}  = 1 - h (\omega t),\;g_{22}  = 1 + h (\omega t),\;g_{33}  = 1$ 
with $h (\omega t)= \sin ^2 \left( \theta  \right)\sin \left( {2\psi } \right) h_0 \sin \left( {\omega t }\right)$. 
The angles $\theta$ and $\psi$ represent the orientation of the direction of the gravitational wave with respect 
to the detector axis. These angles take values from  $\theta = 0...\pi$ and $\psi = 0...2\pi$ \cite{Hendry07}. As an example, 
for an UD detector consisting of a particle with a magnetic moment in a uniform magnetic field the angles of orientation, 
$\theta$ and $\psi$, are between the direction of the magnetic field and the direction of travel of the incoming plane wave.
The constant $h_0$ is the dimensionless amplitude of the gravitational wave, which can be written as $h_0 = G \frac{m_0}{c^2 r_0}$ where $m_0$ and $r_0$ are the relevant mass and distance of the system and $G \frac{m_0}{r_0}$ is a potential energy per mass. 
The amplitude squared of the wave is proportional to the energy density \cite{Carrol03}, 
$T_{00}=\frac{1}{32 \pi} \frac{c^2}{G} \omega^2 h_0^2$. The graviton is taken as the quanta 
of the gravitational wave with energy $\hbar \omega$. 

The spectrum, $S(E)$, observed by the UD detector can be viewed as a production rate of photons per unit volume. 
It is given by the product of the density of states, $\rho (E)$, and the response function $F(E)$ 
\cite{BD,letaw, Akhmedov07,Singleton11,rad},

\begin{equation}
\label{spectrum}
\begin{array}{l}
S(E)=n_{general}  - n_{inertial}  = 2\pi  \rho (E)F(E) \\ 
  = 2\pi  \rho (E)\int_{ - \infty }^{ + \infty } {e^{ - i\Delta E \Delta \tau  } } \left( G^+  (s ) - G^+ (\Delta \tau) \right) 
	d (\Delta \tau) . \\ 
 \end{array}
\end{equation}
In this expression for $S(E)$ we have taken $\hbar =1$ and $c=1$. In (\ref{spectrum}) $\Delta E = E_{up} - E_{down}$ 
is the difference between the energies of the upper and lower energy level of the UD detector \cite{Akhmedov07}. 
We will take $E_{down} =0$ so that from now on we can write $\Delta E \rightarrow E_{up} \equiv E$. The number density 
of the inertial detector, $n_{inertial}$, has been subtracted out to eliminate the singularity in the integrand \cite{rad,letaw} of 
equation (\ref{spectrum}). The trajectory for a particle in an inertial frame is $x(\Delta \tau)=(\Delta \tau,0,0,0)$ which
has an associated Wightman function of $G^+ (\Delta \tau)=\frac{1}{4 \pi^2 \Delta \tau^2}$. The particle trajectory in the curved space of a passing gravitational wave is given by $x(\Delta \tau ) =(\gamma \Delta \tau, \Delta x,0,0)$ where 
$\gamma^{-2}=1- \Delta {\dot x}^2$ and $\Delta x$ is the deviation of a particle's trajectory due to the gravitational 
wave background. Substituting this into the general number spectra in equation (\ref{spectrum}) the integral becomes

\begin{equation}
\label{integral}
\begin{array}{l}
F(E, \theta ,\psi) = \int_{ - \infty }^{ + \infty } {e^{ - i E \Delta \tau } } \left( {G^ + (s) - G^ +  (\tau)} \right)
d (\Delta \tau) \\ 
  = \frac{1}{{2\pi ^2 }}\int_0^{ \infty } {\cos \left( {E \Delta \tau} \right)  } \left( {\frac{1}{\gamma^2 \Delta \tau^2  - \Delta x^2} - \frac{1}{{\Delta \tau^2 }}} \right) d (\Delta \tau) . \\ 
 \end{array}
\end{equation}

\noindent Equation (\ref{integral}) is correct in general for strong gravitational waves since up to now we
have placed no restrictions on $h_0$ or $\gamma$. The dependence of $F(E, \theta, \psi )$ on the angles $\theta , \psi$ 
comes from $\Delta x$ as we will see shortly. Without any approximation this $F(E, \theta ,\psi)$ for the plane wave
background, as given in equation (\ref{integral}), must be evaluated numerically.

In order to make an analytical evaluation of $F(E, \theta ,\psi)$ we take the limit of a weak gravitational wave 
with $\gamma \approx 1$ and $h_0 << 1$. To first order the trajectory along a null geodesic \cite{Taylor14} in the $x$ 
direction is $\dot x = \left( {1 + \frac{1}{2} h} \right)$. The separation, $\Delta x$, can be calculated from 
the difference in the null geodesic between particles in the inertial and curved space-times, 
$\Delta x = \left( {1 + \frac{1}{2} h} \right) \Delta \tau  -  \Delta \tau  = \frac{1}{2} h \Delta \tau $. Substituting 
this $\Delta x$ into equation (\ref{integral}) and using $\gamma \approx 1$ and $h_0 << 1$ we obtain

\begin{equation}
\label{common}
\begin{array}{l}
   F(E,\theta ,\psi)
   { = \frac{1}{{8\pi ^2 }}\int_0^{ \infty } {\cos \left( {E \Delta \tau} \right)} 
	{\frac{{h^2 }}{{\Delta \tau ^2 (1  - \frac{1}{4}h^2 ) }}} 
	d(\Delta \tau) \approx \frac{1}{{8\pi ^2 }}\int_0^{ \infty } {\frac{\cos \left( {E \Delta \tau} \right) {h^2 }}{{\Delta \tau ^2 }}} 
	d(\Delta \tau).}  \\
\end{array}
\end{equation}

\noindent With this approximation the RHS of equation (\ref{common}) can be integrated. Recalling that 
$h (\Delta \tau, \theta, \psi )=  \sin ^2 \left( \theta  \right)
\sin \left( {2\psi } \right) h_0 \sin \left( {\omega \Delta \tau }\right)$ (this dependence 
of $h$ on $\theta , \psi$ is the reason $F$ in (\ref{integral}) and (\ref{common}) depends on $\theta , \psi$) we find,

\begin{equation}
\label{aboveE}
\begin{array}{*{20}c}
   {F(E,\theta ,\psi ) = \frac{1}{{64\pi }} \left( {2  \omega  -  E -  
	\frac{{\left( { E - 2 \omega } \right)}}{{\sqrt {\left( { E - 2  \omega } \right)^2 } }} } (2 \omega - E) \right)
	\sin ^4 \left( \theta  \right)\sin ^2 \left( {2\psi } \right)h_0^2 .}  \\
\end{array}
\end{equation}

\noindent Finally we take the integral of (\ref{aboveE}) over all possible orientations of the direction of the 
gravitational wave with the orientation of the UD detector, ({\it i.e.} integrating over $\theta = 0...\pi$ and 
$\psi = 0...2\pi$) to find

\begin{equation}
\label{aboveEcomplete}
\begin{array}{l}
   F(E)   = \frac{3 \pi}{{256 }}\left( {2  \omega  - E} \right)h_0^2 ~~~{\rm for} ~~ E<2 \hbar \omega ~~~;~~~
	{\rm and} ~~~ F(E) = 0 ~~~{\rm for} ~~ E > 2 \hbar \omega.
\end{array}
\end{equation}

\noindent The response function $F(E)$ vanishes for energies above the cutoff $E> 2 \hbar \omega$. Note that this cut-off is 
also the same as that seen in muon decay \cite{hm} \cite{griffiths}. The meaning of the cut-off 
in (\ref{aboveEcomplete}) can be seen to have a similar kinematic origin -- for the processes $G + G \rightarrow \gamma + \gamma$
and $G \rightarrow G + \gamma + \gamma$ there is at most $2 \hbar \omega$ of energy available (each graviton contributes
$\hbar \omega$ of energy). Thus for $E > 2 \hbar \omega$ neither process can proceed energetically.

The density of states in equation (\ref{spectrum}) is $\rho(E)=\frac{{E}^2}{2 \pi^2}$ \cite{letaw}. 
Using this along with the response function in equation  (\ref{aboveEcomplete}), and restoring the factors of
$\hbar$ and $c$, gives the spectra detected by the UD detector

\begin{equation}
\label{radiationspectra}
\begin{array}{l}
   S(E)   = \frac{3 }{{512 \pi  \hbar^3 c^3}}{E}^2\left( {2 \hbar \omega  - E} \right) h_0^2 ~~~{\rm for} ~~ E<2 \hbar \omega ~~~;~~~
	{\rm and} ~~~ S(E) = 0 ~~~{\rm for} ~~ E >2 \hbar \omega.
\end{array}
\end{equation}

\noindent The result in (\ref{radiationspectra}) gives the spectra of the UD detector $S(E)$. Equation 
(\ref{radiationspectra}) is the main result we want to emphasize in this essay. 
As previously mentioned $S(E)$ has the same form as certain Standard Model processes 
(see for example \cite{hm} \cite{griffiths}). The functional form of $S(E)$ in equation (\ref{radiationspectra}) is a 
$Beta(3,2)$ distribution and such distributions are common for particle decay processes. In Feynman diagram language, 
the processes implied by $S(E)$ from (\ref{radiationspectra}) would be $G + G \rightarrow \gamma + \gamma$ or 
$G \rightarrow G + \gamma + \gamma$ where the gravitons take the place of the muon and electron and the two photons take the places of the neutrinos. 
For $G + G \rightarrow \gamma + \gamma$ the gravitons are completely transformed into photons while for 
$G \rightarrow G + \gamma + \gamma$ the initial graviton is transformed into a lower energy graviton.  
For either process energy is transferred completely or partly into photons and the gravitational wave
will be attenuated.  

Recalling that $S(E)$ from (\ref{radiationspectra}) can be viewed as a production rate of photons per unit volume
for the UD detector we can use it to calculate the decay/attenuation of gravitational waves via the above
processes. By energy conservation the photon production spectrum for the UD detector will come from the gravitons. The ``decay"
rate for gravitons per unit volume is then given by,

\begin{equation}
\label{diffGamma}
\hbar  \frac{d\Gamma}{dV}  = \int\limits_0^{2\hbar \omega } {S(E) dE}.
\end{equation}

The production rate is a function of the dimensionless gravitational wave amplitude, $h_0$, but the physical 
amplitude falls off as $r^{-1}$ for an expanding gravitational wave. The dimensionless gravitational wave amplitude can 
be written as $h_0  = G\frac{{m_0 }}{{c^2 r_0 }}$ where $m_0$ and $r_0$ are an effective mass and distance respectively. 
The physical amplitude for an expanding gravitational wave can then be written as $h_r  = \frac{{\xi _0 }}{r}$ where 
$\xi _0= \frac{{G m_0 }}{{c^2 }}$, which has dimensions of length. Replacing $h_0$ with $h_r$ in equation 
(\ref{diffGamma}) and integrating over all energies in the spectra yields,

\begin{equation}
\label{diffGamma2}
d\Gamma  = \frac{1}{{128\pi c^3 }}\frac{{\xi _0^2 }}{{r^2 }}\omega ^4 dV = \frac{{\xi _0^2 }}{{32c^3 }}\omega ^4 dR.
\end{equation}

\noindent One can immediately integrate the RHS of equation (\ref{diffGamma2}) from $R=0$ to some radius $R=r$ to obtain a decay rate of $\Gamma = \frac{{\xi _0^2 }}{{32c^3 }}\omega ^4 r$. The four factors of $\hbar$ coming from $\hbar \omega$ are 
canceled by the three factors of $\hbar$ coming from $S(E)$ and the one factor coming from the LHS of equation (\ref{diffGamma}).

The decay of gravitons can be expressed in terms of a mean path length, $\left\langle r \right\rangle  = c\left\langle \tau  \right\rangle  = c \Gamma ^{ - 1}$. Integrating equation (\ref{diffGamma2}) from $0$ to $\left\langle r \right\rangle$ and substituting $\Gamma = \frac{{\xi _0^2 }}{{32c^3 }}\omega ^4 r$ in the  mean path length expression gives,

\begin{equation}
\label{pathlength}
\left\langle r \right\rangle  = \frac{c}{\Gamma } = \frac{{32c^4 }}{{\xi _0^2 \omega ^4 \left\langle r \right\rangle }} \to \left\langle r \right\rangle  = \frac{{2\sqrt 8 c^2 }}{{\xi _0 \omega ^2 }}.
\end{equation}

\noindent Due to the inverse relation between the mean free path, $\left\langle r \right\rangle$, and $\xi _0$ and 
$\omega ^2$, gravitational waves are effected less, and therefore attenuated less, by this conversion of gravitons 
into photons for  low frequency ($\omega \ll 1$) and small amplitudes ($\xi_0 \ll 1$). This is similar to the conclusion 
reached by Mondanese \cite{Mondanese95} using ``general kinematical arguments''. For a gravitational wave source on the order 
of a Solar mass, $m_0 \sim 2 \times 10^{30}$ kg, we find $\xi _0  \sim  10^3 m$. Next assuming a gravitational wave frequency 
of $\omega \sim 10^{2}s^{-1}$, the approximate mean decay length would be 
$\left\langle r \right\rangle  \sim 10^{11} m = 10^{-5} ly$. This is a short attenuation distance especially 
considering that the decay rate, $\Gamma$, is normally relatively low. However, this short attenuation length 
is not completely surprising since a small decay rate can be compensated for if it acts over a large distance 
traveled by the gravitational wave. Also for strong gravitational wave sources with large amplitudes, $\xi_0$, and
large frequencies, $\omega$, the decay rate, $\Gamma = \frac{{\xi _0^2 }}{{32c^3 }}\omega ^4 r$, becomes larger. 
Thus for strong gravitational waves one would need to numerically integrate (\ref{integral}) without the approximations 
and then repeat the above development. 

The mean free path for graviton decay in equation (\ref{pathlength}) is based on a semi-classical calculation with the gravitational wave treated classically and the vacuum quantum mechanically. The quantum nature for the graviton appears here only via 
the particle energy, $\hbar \omega$, and through the cutoff in the energy distribution, $2 \hbar \omega$. It would be interesting to compare this result to a more complete quantum field theory calculation of graviton decay following DeWitt \cite{DeWitt67} 
or Bjerrum-Bohr {\it et al.} \cite{Bohr08}. While such calculations have the advantage of including spin and more carefully taking account of energy-momentum conservation, any mean free path calculation for graviton transition rates would include a similar calculation over the volume of space as in equation (\ref{diffGamma2}). The UD detector calculation, followed by the mean 
free path estimate given above, indicates that detecting gravitational waves at large distances may be hard or impossible
due to the attenuation of the gravitational wave coming from the conversion of gravitons to photons.

\begin{acknowledgments}
DS acknowledges support by a Grant (number 1626/GF3) in Fundamental Research in Natural Sciences by the Science Committee of the Ministry of Education and Science of Kazakhstan.
\end{acknowledgments}


\begin{thebibliography}{99}

\bibitem{BD} N.D. Birrell and P.C.W. Davies, {\it Quantum fields in curved space},
(Cambridge University Press, Cambridge 1982).

\bibitem{Parker09} Leonard E. Parker and David J. Toms, {\it Quantum field theory in curved space}, 
(Cambridge University Press, Cambridge 2009).

\bibitem{DeWitt75} Bryce S. DeWitt, {\it Quantum field theory in curved spacetime}, (North Holland Publishing, Amsterdam 1975), 
Physics Reports {\bf 19}, 295-397 (1975).

\bibitem{Hawking74} S.~W.~Hawking, Commun. Math. Phys.  {\bf 43}, 199 (1975) [Erratum-ibid.\  {\bf 46}, 206 (1976)].

\bibitem{Unruh76} W. G. Unruh, Phys. Rev. D {\bf 14}, 870 (2001).

\bibitem{Bell83} J. S. Bell and J. M. Leinaas, Nucl.Phys. B {\bf 212}, 131 (1983).

\bibitem{letaw} J. R. Letaw and J. D. Pfautseh, Phys. Rev. D {\bf 22} 1345 (1980); 
J.R. Letaw Phys. Rev. D {\bf 23}, 1709 (1981).

\bibitem{Akhmedov07} E. T. Akhmedov and D. Singleton,  Int. J. Mod. Phys. A {\bf 22} 4797 (2007).

\bibitem{Akhmedov07a} E. T. Akhmedov and D. Singleton,  Pisma Zh. Eksp. Teor. Fiz. {\bf 86}, 702 (2007).  [JETP Letts. {\bf 86}, 615 (2007)].

\bibitem{Kallberg06} A. K{\"a}llberg, G. Brodin, M. Marklund, Class. Quant. Grav. {\bf 23}, ``Photon-graviton pair conversion", L7-L13 (2006).

\bibitem{Brodin06} G. Brodin, D. Eriksson, M. Marklund, Phys. Rev. D {\bf 74} 124028 (2006).

\bibitem{Bohr08} N. E. J. Bjerrum-Bohr, Barry R. Holstein, Ludovic Plant, Pierre Vanhove, IPhT/t14/148, IHES/P/14/32, ``Graviton-Photon Scattering", arXiv:1410.4148  [gr-qc].

\bibitem{Mondanese95} G. Modanese, Phys. Lett. B {\bf 348}, 51-54 (1995).

\bibitem{Taylor14} Edmund Bertschinger and Edwin F. Taylor ``Chapter 16, Gravitational Waves" (2014), www.eftaylor.com/exploringblackholes/, GravWaves141128v2.

\bibitem{Hendry07} M. Hendry, ``An Introduction to General Relativity, Gravitational Waves and Detection Principles", Second VESF School on Gravitational Waves (2007).

\bibitem{Carrol03} S. Carrol, ``Spacetime and Geometry: An Introduction to General Relativity"  (Addison-Wesley, 2003) pg. 311.

\bibitem{Singleton11} Douglas Singleton and Steve Wilburn, Phys. Rev. Lett. {\bf 107}, 081102 (2011);  arXiv:1110.1099 [hep-th].

\bibitem{rad} N. Rad and D. Singleton, Eur. Phys. J. D {\bf 66}, 258 (2012);  arXiv:1110.1099 [hep-th].

\bibitem{hm} F. Halzen and A.D. Martin, {\it Quarks \& Leptons: An Introductory Course in Modern Particle Physics}, pg. 263
(John Wiley \& Sons, 1984).

\bibitem{griffiths} D. Griffiths, {\it Introduction to Elementary Particles} 2$^{nd}$ edition, pg. 313 (Wiley-VCH, Weinheim, 2008).

\bibitem{DeWitt67} Bryce S. DeWitt, Phys. Rev. {\bf 162}, 1239-1256 (1967).

\end{thebibliography}
\end{document}